\documentclass[12pt]{article}
\usepackage{amsmath}
\usepackage{graphicx}
\date{}
\begin{document}
\title{The onset of the bipolar flavor conversion of supernova neutrinos}
\author{{\bf Indranath Bhattacharyya}\vspace{0.2cm}\\
Department of Mathematics\\
Acharya Prafulla Chandra Roy Government College\\Himachal Vihar, Matigara, Siliguri-734010, West Bengal
\\E-mail :
$i_{-}bhattacharyya@hotmail.com$\\}\vskip .1in
\maketitle
\begin{abstract}
The study of supernova neutrinos result an interesting non-linear phenomenon, consisting
of three phases: synchronized oscillation phase, bipolar flavor conversion phase and the
phase of spectral split. In the collective oscillation of supernova neutrino the self energy is
not a constant but varies adiabatically, which is responsible to have such different phases.
In this article the transition point from synchronized oscillation to bipolar phase is studied
numerically as well as analytically. The numerical results yielding different graphs depending
on different values of possible small but non-vanishing mixing angles show the onset of the
bipolar phase from the synchronized phase varies as the mixing angle. But the analytical
study in terms of a spinning top model results a unique onset condition, which is independent
of the choice of mixing angle. Such discrepancy between numerical results and analytical
results is explained properly.
\vspace{0.5cm}
\vspace{0.2cm}\\PACs: 14.60.Pq; 97.60.Bw
\end{abstract}
\section{Introduction:}
Supernova neutrino study results the interesting non-linear collective oscillation phenomena of neutrinos \cite{Amol}. In the very earlier calculations incorporating the refractive index of
the neutrino-neutrino forward scattering, the off diagonal
refractive indices representing the flavor transitions of
neutrinos had been left out. It was, later,
rectified by Pantaleone\cite{Pantaleone}. The core-collapse
supernova shows that neutrinos can also have a nontrivial
background to themselves when their density is extremely high. In
a SN envelop MSW effect comes into effect when $\lambda$ is close
to $\omega$. For a few seconds after burst, i.e., when $r>100$
$km$ (large radii) the $\lambda$ is close to $\omega$, but when
$r\ll O(100)$ $km$ then $\lambda\gg\omega$. That does not imply
the flavor transitions get suppressed, rather at small $r$
neutrino and antineutrino densities are high enough to make the
self-interaction effect important, which leads to the collective
oscillations phenomena of neutrinos. Such effect that is evident when the number density of the neutrinos is very high, especially during supernova neutrino emission, was observed and studied by
a number of researchers and scientists \cite{Pantaleone, Samuel1993,
Samuel1996, Qian1995, Qian2006, Raffelt1, Raffelt2, Duan1, Duan2,
Duan3, Raffelt3, Mirizzi}. Although in the core collapse
supernova the collective effect plays a significant role, but the
role of ordinary MSW effect is also important \cite{Amol1}. Neutrino
flavor oscillation is thus a complicated multi-scale dynamical
problem which involves both collective and MSW effect. The existence of non-linear self interaction effect could be realized when one can see one \cite{Duan2, Raffelt4} or more \cite{Basu1} splits in both $e$-type or $x= (\mu$ or $\tau)$-type of neutrinos in the spectrum. This is the ultimate fate of the collective oscillations scenario, although it goes through another two phases - bipolar and synchronized phase. In the numerical simulation the oscillation ranges of
different self-interaction effects have been observed as follows \cite{Mirizzi}: \vspace{0.2cm}\\
$\mu\geq\mu_{sup}$: synchronized
oscillation,\hspace{0.2cm}$(r\leq 55$ km)\\
$\mu_{inf}\leq\mu\leq\mu_{sup}$: bipolar
oscillation,\hspace{0.2cm}$(55$ km $\leq$ $r\leq 100$ km)\\
$\mu\leq\mu_{inf}$: spectral split,\hspace{0.2cm}$(r\geq 100$ km)\vspace{0.2cm}\\
The analysis has been done with the mixing of two flavors, although some extensive works \cite{Basu2, Basu3} have already been carried out with taking all three types of neutrinos.
It is well known that in the first phase of the collective oscillations all the
neutrino and anti-neutrino modes oscillate with a single
synchronized frequency. In the spin-precession analogy the flavor
polarization vectors of all neutrino and anti-neutrino modes form
one big internal magnetic field that precesses in the external magnetic field of mass basis. To begin the collective oscillations it is mandatory that such internal magnetic field is much stronger than the external magnetic field formed in the flavor space by the
strong neutrino-neutrino interaction. Such picture may be
visualized as the motion of a spherical pendulum in the flavor
space having certain frequency. When the oscillation enters
into the bipolar phase the pendulum in the flavor space starts
nutating. For inverted mass hierarchy a small
mixing angle can make a complete conversion of neutrinos
(antineutrinos) to the other flavor provided the case is completely symmetric in terms of lepton number. The lepton asymmetry generates a spin and as a result complete conversion is not possible. In the pendulum analogy it goes to the southern latitude but cannot go completely to the south pole, particularly in the initial stage of bipolar oscillation. Thus in that phase the nutation as well as precession occurs simultaneously.\\\indent In this paper we are going to study the onset of the bipolar phase from the synchronized phase. In other words we would like to study how the onset point depends on different parameters. We are strongly motivated by a partial study of this problem by Hannested et al. \cite{Raffelt3} and Duan et al. \cite{Duan2}. The picture can be realized as the motion of a spinning top. If one can study such motion it is observed that initially top executes a pure precession and at certain point it starts to nutate upto a southern latitude. We are at the point of study of such onset. It is to be noted that in our entire literature we have used the notation $\theta_{0}$ as the twice the vacuum mixing angle. In the section-2 we have taken the equation of motion of the collective oscillation phenomenon and shown numerically how the onset is attained with the changing self-interaction energy $\mu$. In the next section we have considered the dynamics of spinning top and deduced the onset condition in the top language. Here we have also studied analytically how the onset condition depends on the initial inclination of the top. That leads to a relation between the transition point and the vacuum mixing angle of the neutrino oscillation. In the section-4 we have drawn a direct analogy between spinning top mechanism and the collective oscillation picture and translate the onset condition in terms of $\mu$ and $\omega$. In this section we have obtained numerically the relation between onset point and the vacuum mixing angle. The values of all parameters obtained analytically are fixed by the numerical simulation.
\section{Basic equations of motion and the onset}
The collective oscillation phenomena starts with synchronized oscillations in which all modes of neutrinos oscillate with a single frequency. No significance conversion occurs in this phase. The conversion is significant when it enters into its bipolar phase. In the inverted hierarchy at the end of synchronized phase the nature of the oscillation suddenly becomes exponential. That is called onset, a transition point from synchronized to bipolar phase. In the study of this onset let us consider, for simplicity, a single mode of neutrino and antineutrino. In absence of matter the corresponding equations of motion of the collective oscillation are given by
\begin{equation}
\partial_{t}\mathbf{P}=\omega(\mathbf{B}\times\mathbf{P})+
\mu(\mathbf{D}\times\mathbf{P})
\end{equation}
\begin{equation}
\partial_{t}\mathbf{\overline{P}}=-\omega(\mathbf{B}\times\mathbf{\overline{P}})+
\mu(\mathbf{D}\times\mathbf{\overline{P}})
\end{equation}
where,
$\mathbf{P}$ and $\mathbf{\overline{P}}$ are the corresponding polarization vectors in flavor space for neutrino and antineutrino respectively. Those are completely asymmetric in nature and thus their $z$-component cannot develop identically. We have considered the inverted hierarchy and kept the vacuum frequency $\omega$ always positive. This is a simplified model of the collective oscillation, but good to study the onset condition. In the collective picture the self energy $\mu$ is not a constant but varies adiabatically. We shall discuss its adiabaticity later on. For a large $\mu$ the oscillation type is synchronized, whereas for a small $\mu$ a clear bipolar nature is observed. This can be well visualized from the Figure-1, where the evolution of $P_{z}$ is shown for a small $\mu$ (=10) as well as large $\mu$ (=200). Now, one can write the above equations in terms of $\mathbf{S}=\mathbf{P}+\mathbf{\overline{P}}$ and $\mathbf{D}=\mathbf{P}-\mathbf{\overline{P}}$ as follows:
\begin{equation}
\partial_{t}\mathbf{S}=\omega(\mathbf{B}\times\mathbf{D})+
\mu(\mathbf{D}\times\mathbf{S})
\end{equation}
\begin{equation}
\partial_{t}\mathbf{D}=\omega(\mathbf{B}\times\mathbf{S})
\end{equation}
It is now interesting study the self interaction effect. Due to very high value of $\mu$ relative to $\omega$ the polarizations vectors $\mathbf{P}$ and $\mathbf{\overline{P}}$ along with their prcessions about $\mathbf{B}$ also start to precess around each other with a very small angle. The situation can be thought as both of the polarization vectors precess around the external magnetic field $\mathbf{D}$ as well as around an internal magnetic field $\mathbf{D}$. As a result a peculiar non-linear effect is developed. Let us consider a very small angle between $\mathbf{D}$ and $\mathbf{S}$ as $\delta$. Then from equation (4) it is obtained
$$
\partial_{t}\mathbf{D}=\omega\frac{S}{D}[\mathbf{B}\times(\mathbf{D}+\mathbf{D}_{\perp}\delta)]
$$
It shows that for a small $\delta$ we can neglect the second term. It leads to the situation that the internal magnetic field precesses around the external magnetic field with the synchronized frequency $\omega\frac{S}{D}\approx\omega\frac{1+\alpha}{1-\alpha}$, where $\mathbf{\overline{P}}(0)=\alpha\mathbf{P}(0)$. We would like to examine the situation in a more precise manner.\\\indent Let us construct a vector $\mathbf{Q}$ as
\begin{equation}
\mathbf{Q}=\mathbf{S}-\frac{\omega}{\mu}\mathbf{B}
\end{equation}
It can easily be calculated
\begin{equation}
Q=[(1+\alpha)^{2}+(\frac{\omega}{\mu})^{2}-2\alpha\frac{\omega}{\mu}\cos\theta_{0}]^{\frac{1}{2}}\approx(1+\alpha)
\end{equation}
For a large $\mu$ the vector $\mathbf{Q}$ is approximately equal to $\mathbf{S}$.
Now using the equation (3) one can express the internal magnetic field as
\begin{equation}
\mathbf{D}=\frac{1}{2\mu}({\mathbf{q}}\times\dot{{\mathbf{q}}})+\frac{\mu}{2}\sigma\mathbf{q}
\end{equation}
where, $q$ is the unit vector along $\mathbf{Q}$. Clearly, the above expression is equivalent to the angular momentum. The first term in the right hand side represents the orbital angular momentum, whereas the second term stands for the spin angular momentum with
$$\sigma=\mathbf{D}.\mathbf{q}\approx 1-\alpha$$
It is now easy to express the equation of motion in terms of $\mathbf{Q}$. An important point is to be noted that to construct such equation we have neglected the term $\dot{\mu}$ as it is very small, but of course not zero. Now the corresponding equation of motion is
\begin{equation}
\frac{1}{\mu}({\mathbf{q}}\times\ddot{{\mathbf{q}}})+\sigma \dot{{\mathbf{q}}}=\omega Q({\mathbf {B}}\times {\mathbf{q}})
\end{equation}
For sufficiently large $\mu$ the first term of the equation can be dropped and the resulting equation clearly shows that $\mathbf{q}$ precesses round $\mathbf{B}$ with synchronized frequency. If the case is symmetric, then $\sigma=0$ and the above equation shows there will be an oscillation, called nutation. In general, there must be a combination of precession and nutation. But the question is when such bipolar motion starts.\\\indent To study the such transition point numerically we must consider that the $\mu$ varies very slowly with time. In the numerical calculation of Hannestead et al. \cite{Raffelt3} the adiabaticity of $\mu$ is achieved by considering its expression as the function of $t^{-4}$. In our numerical calculation we take $\mu=\mu_{I}e^{-kt}$ \cite{Raffelt4}, where $k$ is the adiabatic parameter having the dimension of inverse time, although it cannot be so small that $\mu$ remains practically a constant. On the other hand if $k$ becomes large then $\mu$ falls rapidly, which violates the adiabacity. Therefore, it is important to fix $k$ properly. Its extent of smallness describes how slow $\mu$ decreases. In our numerical calculations we have considered $k=0.1\omega$. Now, in the Figure-2 the evolution of $P_{z}$ and $\bar{P}_{z}$ are plotted against $(\frac{\mu}{\omega})^{\frac{1}{2}}$ with three different vacuum mixing angles. A sharp deviation of $P_{z}$ from its initial value denotes the onset. It is observed that the onset decreases with decreasing the mixing angle. That indicates the longer time is required to achieve the onset with smaller mixing angle. The figure is very much similar to that of Hannestad et al. \cite{Raffelt4} \footnote {Note that in \cite{Raffelt4} the relative flux were plotted, but we have plotted here $P_{z}$ and $\bar{P}_{z}$.}. It was remarked in this paper that the onset value of $(\frac{\mu}{\omega})^{\frac{1}{2}}$ would  decrease logarithmically with decreasing small mixing angle. We shall examine that case both analytically as well as numerically.
\section{Onset of a spinning top}
In this section we study the dynamics of a spinning top and find under what condition it slips. If we closely looks at the top we see it precesses with certain frequency. At some point of time it slips and start nutating. The precession and nutation continues and ultimately it falls down. But we are interested about the point at which the nutation is started. The Lagrangian of the top equation is given by \cite{Goldstone}
\begin{equation}
L=\frac{I_{1}}{2}(\dot{\theta}^{2}+\dot{\phi}^{2}\sin^{2}\theta)+\frac{I_{3}}{
2
}(\dot{\psi}+\dot{\phi}\cos\theta)^{2}-mgl\cos\theta
\end{equation}
where, $\theta$ is a small inclination of the axis of the top to the vertical. The top is here spinning about its axis. $\phi$ and $\psi$ denote the azimuth of its rotation around vertical and spin axis respectively. $I_{3}$ represents the component of the moment of inertia along its spin axis, whereas $I_{1}$ is that along a line lying in the plane perpendicular to the spin axis, also called figure axis. The momentum conjugate to a rotation angle is the component of the total angular momentum along the axis of rotation. In that case the torque (due to gravity) is along the line of nodes (the intersecting line between the horizontal plane and the plane perpendicular to the top). Therefore, there is no torque either along the spin axis or along the vertical axis as both of them are perpendicular to the line of nodes. Hence the components of the angular momentum along these two axes ($\psi$ and $\phi$) must be constant in time. Thus we get
\begin{equation}
 p_{\psi}=\frac{\partial
L}{\partial\dot{\psi}}=I_{3}(\dot{\psi}+\dot{\phi}\cos\theta)=constant=I_{1}a
\end{equation}
\begin{equation}
p_{\phi}=\frac{\partial
L}{\partial\dot{\psi}}=(I_{1}\sin^{2}\theta+I_{3}\cos^{2}{\theta})\dot{\phi}+I_{
3} \dot{\psi}\cos\theta=constant=I_{1}b
\end{equation}
We are now in the situation when there is a pure precessional mode in the sense that there is no nutation, but there must be a spin around the figure axis. Now using the above two equations we can get the expression of the energy in terms of $\theta$ equation as
\begin{equation}
E_{c}=\frac{I_{1}\dot{\theta}^{2}}{2}+\frac{I_{1}}{2}\frac{(b-a\cos\theta)^{2}}{
\sin^{2}\theta}+mgl\cos\theta
\end{equation}
Note that in the above expression the term $\frac{I_{3}}{2}(\dot{\psi}+\dot{\phi}\cos\theta)^{2}$, which is essentially a constant of motion, is absorbed. It clearly shows an expression of energy to an one-dimensional problem with effective
K.E. term $\frac{I_{1}\dot{\theta}^{2}}{2}$ and potential term $\frac{I_{1}}{2}\frac{(b-a\cos\theta)^{2}}{\sin^{2}\theta}+mgl\cos\theta$. Now choosing $u=\cos\theta$ we can obtain an expression as follows:
\begin{equation}
f(u)=\dot{u}^{2}=(1-u^{2})(\alpha-\beta u)-(b-au)^{2}
\end{equation}
where,
\begin{equation}
\alpha=\frac{2E_{c}}{I_{1}},\hspace{0.2cm}\beta=\frac{2mgl}{I_{1}},\hspace{0.2cm
}u=\cos\theta
\end{equation}
The physical motion of the top can occur only when $f(u)$ is positive. For,
$f(u)\leq0$ there will be no nutation at all. Thus the nutation will start only
when $f(u)>0$. It is now observed that the dominant term in $f(u)$ is $\beta u^{3}$. This term must be positive for the large positive value of $u$, whereas negative for $u=1$, except for the case when $u=1$ itself is a root of the equation. Hence abreast one root must lie in the region $u>1$. But in the case when $u=1$ is a root of the equation all three roots may lie in $(-1,1)$, the region corresponding to the real angle. Now in general, if all three roots are real then one lies outside of $(-1,1)$, whereas other two lie inside. Let us consider the roots lying inside. Two possibilities are there; either they are equal or unequal. The double root implies there is no positive part of the curve inside $(-1,1)$ which signifies the top continues spinning. Let us consider the situation when the initial point $u=u_{0}$ is a root of the equation $f(u)=0$, where the point $u_{0}$ is close to 1. Accordingly all the constants $a$, $b$, $\alpha$ and $\beta$ are fine tuned. Therefore, we can say the dynamics of the top depends on the initial condition. Now, if $u_{0}$ is the double root the third one cannot enter into the interval $(-1,1)$ and the top merely continues its spin. The situation can be well explained by the Figure-3. The dark dashed line represents the situation where the root $u_{0}$ lying inside $(-1,1)$ is a double root. We see that there is no positive part of the curve in $(-1,1)$, only it touches the $u$-axis at $u=u_{0}$, which now becomes double root. The nutation will be possible only when $u=u_{0}$ will be no longer a double root and the top nutates between two roots of $f(u)=0$ in $(-1,1)$. The light dashed line shows that one root comes inside the region $(-1,1)$ (close to -1) and the nutation occurs between $u_{0}$ and that root. We are now going to consider under which circumstances the double root conditions get violated. Using the conditions $f(u)=0$ and $\frac{df}{du}=0$ at $u=u_{0}$ one can easily obtain
\begin{equation}
\cos\theta_{0}\dot{\phi}^{2}-a\dot{\phi}+\frac{\beta}{2}=0
\end{equation}
where the $\dot{\phi}$ is taken at $u=u_{0}$. The above equation will be valid only when $\frac{a^{2}}{2\beta}\geq\cos\theta_{0}$. The top precesses so far this condition holds. When $\frac{a^{2}}{2\beta}<\cos\theta_{0}$ the above equation will have no real solution and $u_{0}$ will be no longer a double root. Thus the nutation will start at
\begin{equation}
\frac{a^{2}}{2\beta}=\cos\theta_{0}\approx 1
\end{equation}
whereas,
\begin{equation}
\dot{\phi}=\frac{a}{2\cos\theta_{0}}\approx\frac{a}{2}
\end{equation}
The equation (16) represents the onset condition and the equation (17) gives the value of $\dot{\phi}$ at the onset point. Thus the onset condition is deduced. It is observed that the onset depends on the initial inclination $\theta_{0}$. But such $\theta_{0}$ dependence has negligible effect as $\cos\theta_{0}$ is close to 1. The significant $\theta_{0}$ dependence of the onset value, which is observed from Figure-2, can be realized in the other way. We shall discuss it in the next chapter.
\section{Onset at different $\theta_{0}$}
We intend to draw the analogy of this collective oscillation with the spinning top picture. The motivation is to study the transition from synchronized oscillation (pure precession mode) to bipolar oscillation (precession plus nutation). With the aid of equation (7) the Lagrangian of the collective oscillation can be written as \cite{Raffelt3, Mirizzi}
\begin{equation}
L=\frac{\dot{\mathbf{q}}^{2}}{2\mu}+\frac{\mu\sigma^{2}}{2} -\omega Q\cos\theta
\end{equation}
Let us now compare the various terms as follows \cite{Mirizzi}:
\begin{center}
$\mathbf{D}\rightarrow$ angular momentum\\\smallskip $q=l=1$;\hspace{0.5cm} $m l^{2}=\frac{1}{\mu}$;\hspace{0.5cm} $I_{1}=I_{3}=\mu^{-1}$\\\smallskip $\mathbf{B}\rightarrow$ positive vertical axis;\hspace{1cm}$\mathbf{q}\rightarrow$ positive spin axis\\\smallskip $mgl=\omega Q\rightarrow$ gravitational energy
\end{center}
With this analogy it can now easily be found
\begin{equation}
a=\mu\sigma=\dot{\psi}+\dot{\phi}\cos\theta_{0};\hspace{2cm}
\beta=2\omega Q\mu
\end{equation}
Thus the onset condition deduced in the equation (16) takes the form
\begin{equation}
\frac{\mu}{\omega}\approx 4\frac{Q}{\sigma^{2}}
\end{equation}
for small $\theta_{0}$.\\\indent
The same relation has also been deduced by Hannested et al. \cite{Raffelt3}, but the numerical results give a different picture. The Figure-2 shows clearly that as the mixing angle decreases the onset value shifts towards the right, that means onset $\mu$ should decrease subsequently. We have calculated numerically how the onset $\mu$ depends on the mixing angle $\theta_{0}$. In Figure-4 we have plotted the onset $(\frac{\mu}{\omega})^{\frac{1}{2}}$ against $\theta_{0}$, which shows clearly the onset $\mu$ decreases with smaller $\theta_{0}$. In other words the onset $\mu$ depends on the smallness of the mixing angle and Hannestead et al. \cite{Raffelt3} demanded such dependence is logarithmic. But the equation (20) does not show such effect. That discrepancy between theory and the numerical result, that is also present in the paper of Hannestead et al. \cite{Raffelt3}, should be explained properly.\\\indent Let us now recall the top picture. It is observed that if the angle of inclination is made smaller the top will take longer time to reach the onset of the nutation. Such observation is based on noting down the sharp descent from the initial position at which the top was in the mode of pure precession, although it is difficult to locate that onset point accurately. It is shown in the Appendix that for small variation of $\theta$ the $\dot{\phi}$ does not deviate much from $\frac{a}{2}$. Therefore, it can be said even after the onset is achieved the top starts to deviate from its inclination very slowly so that it is very difficult to locate the actual point at which the deviation begins, i.e., the true onset point. The same difficulty arises also in the case of collective oscillation picture. Let us consider $\mu_{0}$ as the value of $\mu$ for which the common onset condition, defined in the equation (20), holds. Now for the small variation $\theta$ the expression $\mu=\mu_{0}e^{-kt}$ takes the form as follows
\begin{equation}
\mu=\mu_{0}(1-kt)
\end{equation}
Taking $q=(\sin\theta\cos\phi, \sin\theta\sin\phi, \cos\theta)$ in the inverted hierarchical scenario the Lagrangian, given by (18) can be written as
\begin{equation}
L=\frac{1}{2\mu}(\dot{\theta}^{2}+\dot{\phi}^{2}\theta^{2})+\frac{1}{2\mu}(\dot{\psi}+\dot{\phi})^{2}-\omega Q\mu
\end{equation}
assuming for a small variation of $\theta$. The smallness of $\theta$ is constrained by
\begin{equation}
\sin\theta\approx\theta;\hspace{2cm}\cos\theta\approx 1
\end{equation}
Therefore such variation is very small. It can be proved (Appendix) that for small variation of $\theta$ the precession angular frequency $\dot{\phi}$ is confined to $\frac{a}{2}=\frac{\sigma\mu}{2}$. Since the change of $\mu$ is adiabatic therefore it does not affect the dynamics of the small variation of $\theta$. It can be understood in the other way. The equation (15) corresponds to the double root condition. It has already been stated that when the double root condition is violated this equation will no longer have the real solution. In terms of collective oscillation the equation has the solution as follows.
\begin{equation}
\dot{\phi}=\frac{\mu\sigma}{2}-\frac{(\mu^{2}\sigma^{2}-4\omega Q\mu)^{\frac{1}{2}}}{2}
\end{equation}
Note that the negative sign of the solution is taken instead of the positive sign. The positive sign implies that initially $\dot{\phi}$ has the value $\mu\sigma$ and finally it comes down to $\frac{\mu\sigma}{2}$. But for negative sign $\dot{\phi}$ starts with $\frac{\omega Q}{\sigma}$ and reaches to $\frac{\mu\sigma}{2}$ in the synchronized regime. Therefore, the second one is consistent with the collective oscillations scenario as initially $\mathbf{q}$ has a synchronized frequency $\frac{\omega Q}{\sigma}\approx\omega\frac{1+\alpha}{1-\alpha}$.\\\indent
Now the significant decrease of $\mu$ could make the term in the square root bracket negative and consequently the double root condition would be violated, as a result the bipolar motion must be started immediately. Therefore, $\dot{\phi}$ remains at $\frac{\mu_{0}\sigma}{2}$ for the small variation of $\theta$.
Now from the Lagrangian given by the equation (22) we can have the linear $\theta$ equation of motion as follows.
\begin{equation}
\ddot{\theta}=\frac{\mu_{0}^{2}\sigma^{2}}{4}\theta
\end{equation}
Solving it with the initial conditions $\theta(0)=\theta_{0}$ and $\dot{\theta}(0)=0$ one can obtain
\begin{equation}
\theta=\theta_{0}\cosh(\frac{\mu_{0}\sigma t}{2})
\end{equation}
In Figure-5 $\theta$ is plotted against $t$ at different $\theta_{0}$'s to yield a family of curves. Each curve exhibits the exponential nature. The line $\theta=0.01$ cuts all the curves at different points, which signifies that in the top picture $\theta=0.01$ will be attained at different time if the initial $\theta_{0}$ are different. In the same manner to attain any other point, say $\theta=0.08$ the time taken by the top will be different if started from different $\theta_{0}$. Therefore, to reach at any particular $\theta$ the time taken must be different for different initial $\theta_{0}$, but the changes of such time scale will be same for all $\theta$ in the given range. It is quite obvious that $\mu$ must change its value from $\mu_{0}$ whenever the angle is changed from initial $\theta_{0}$ to a particular $\theta$. From the equation (21) and (26) it can be deduced
\begin{equation}
\mu=\mu_{0}-\frac{2k}{\sigma}\cosh^{-1}(\frac{\theta}{\theta_{0}})
\end{equation}
Introducing the dimensionless quantity $x=(\frac{\mu}{\omega})^{\frac{1}{2}}$ it can be obtained
\begin{equation}
x=[x_{0}^{2}-\frac{2k}{\omega\sigma}\cosh^{-1}(\frac{\theta}{\theta_{0}})]^{\frac{1}{2}}
\end{equation}
In Figure-6 $x=(\frac{\mu}{\omega})^{\frac{1}{2}}$ is plotted against $-\log\theta_{0}$ for two different values of $\theta$, which yields a pair of parallel straight lines. Such linear nature of a particular curve signifies that there exists logarithmic relation between $\theta_{0}$ and $(\frac{\mu}{\omega})^{\frac{1}{2}}$ as observed numerically by Hannestad et al. It is to be noted carefully that $(\frac{\mu}{\omega})^{\frac{1}{2}}$ is evaluated at $\theta$, not at $\theta_{0}$. If we consider that the onset occurs at $\theta$, not at $\theta_{0}$ then clearly the $(\frac{\mu}{\omega})^{\frac{1}{2}}$ (at $\theta$) varies with $\theta_{0}$. Therefore, the nature of the curves, representing the variation of the values of $(\frac{\mu}{\omega})^{\frac{1}{2}}$, are same for any $\theta$. Eventually, $(\frac{\mu}{\omega})^{\frac{1}{2}}$ is evaluated at $\theta$ instead of $\theta_{0}$. In principle at $\theta_{0}$, whatever it may be, the value of $(\frac{\mu}{\omega})^{\frac{1}{2}}$ is approximately equal to $\frac{(1+\alpha)^{2}}{1-\alpha}$, $\theta$ represents a pseudo onset, which is essentially not fixed.
\section{Concluding Remarks}
We have studied here the onset of a bipolar oscillation for a simple case of single angle approximations. It has helped us to consider the collective oscillation as a model of spinning top. The spin precession analogy of the spinning top gives an idea how onset of the bipolar motion behaves. Our numerical study reveals that onset value of $\mu$ depends on the vacuum mixing angle; A larger mixing angle causes to attain the onset much earlier. We have tried to find its reason. If we watch an exponential curve in a large scale we find that at the beginning the curve seems to be a constant. But if the curve is observed in a small scale it can be clearly verified that the initial portion, which have been observed as constant in the larger scale, would actually have the exponential nature. It is well observed, both numerically as well as analytically, that just after achieving onset the $\theta$ changes exponentially, although initially such change is not very sharp. Therefore, for a while it seems that the $\theta$ does not change at all, which implies that the bipolar motion has not been started. It follows from the equation (28) that the uncertainty $(\Delta x_{0})^{2}=x_{0}^{2}-x^{2}=\frac{2k}{\omega\sigma}\cosh^{-1}(\frac{\theta}{\theta_{0}})$ vanishes only when the onset is evaluated at the point $\theta_{0}$. But in the numerical calculations it is hardly possible to fix the onset exactly at the true position as  $\dot{\phi}$ remains at $\frac{\mu_{0}\sigma}{2}$ for the small variation of $\theta$ from $\theta_{0}$. As a result the uncertainty $(\Delta x_{0})^{2}$ is nonvanishing and considerable. Thus in the numerical calculations what $(\frac{\mu}{\omega})^{\frac{1}{2}}$ is considered as the onset value, cannot be evaluated at the true onset. It seems then that the onset value changes logarithmically with changing $\theta_{0}$. That is not true and on contrary the true onset value of $(\frac{\mu}{\omega})^{\frac{1}{2}}$ must be fixed for any small $\theta_{0}$.
\section{Appendix}
We are at the situation to consider the variation just after the onset is attained. It is important how far such variation is allowed. Essentially the variation is considered in such a way that the angle of inclination remains small in terms of $\sin\theta\approx\theta$. This is quite obvious as we concentrate at the onset and therefore its neighbourhood should not exceed the considerable limit. Now we are going to prove in the small neighbourhood of the onset $\dot{\phi}$ remains constant. Let us now consider the Lagrangian for the top. From the equations of motions for $\phi$ and $\psi$ yield
$$
\ddot{\phi}\cos\theta+\ddot{\psi}-\dot{\phi}\dot{\theta}\sin\theta=0
\eqno{(A1)}$$
$$
(I_{1}\sin^{2}\theta+I_{3}\cos^{2}\theta)\ddot{\phi}+I_{3}\ddot{\psi}\cos\theta+[2(I_{1}-I_{3})\dot{\phi}\sin\theta\cos\theta-I_{3}\dot{\psi}\sin\theta]\dot{\theta}=0 
\eqno{(A2)}
$$
Note that the Lagrangian is free from $\phi$ and $\psi$ and therefore these two can be considered as the cyclic coordinates. Compairing the above two equations with the aid of equation (10) and $
\sin\theta\approx\theta;$ $\cos\theta\approx 1$ one can find
$$
\ddot{\phi}\theta+2\dot{\phi}\dot{\theta}-a\dot{\theta}=0
\eqno{(A3)}$$
This is esentially a linear first order differential equation of $\dot{\phi}$ and $\theta$ having the form
$$
\frac{d\theta}{d\dot{\phi}}+\frac{\theta}{2\dot{\phi}-a}=0
$$
whose solution is found to be
$$
\theta(2\dot{\phi}-a)^{\frac{1}{2}}=const.
$$
Initially, $\theta=\theta_{0}$ and $\dot{\phi}=\frac{a}{2}$ (onset value, assuming $\cos\theta_{0}\approx 1$)
Therefore, $const.=0$ and it is found
$$
\theta(2\dot{\phi}-a)^{\frac{1}{2}}=0\eqno{(A4)}$$

The above equation states for every small variation of non-zero $\theta$
$\dot{\phi}$ remains at $\frac{a}{2}$. Ofcourse this is a valid argument only for small $\theta$. When the $\theta$ becomes considerably large the $\dot{\phi}$ becomes no longer constant.
\section{Acknowledgement}
I am extremely obliged and thankful to Prof. Amol Dighe, Department of Theoretical
Physics, TaTa Institute of Fundamental Research, Mumbai, India, for his proper guidance
and active cooperation. I must convey my thanks to Prof. G. G. Raffelt, Max Planck
Institute for Physics, Munich, Germany to point out this problem and encourage me to solve
it. My special thank goes to Dr. A. Mirizzi to share some useful tricks and information
towards solving this problem.

\pagebreak
\begin{figure}
\begin{center}
\includegraphics[scale=0.7]{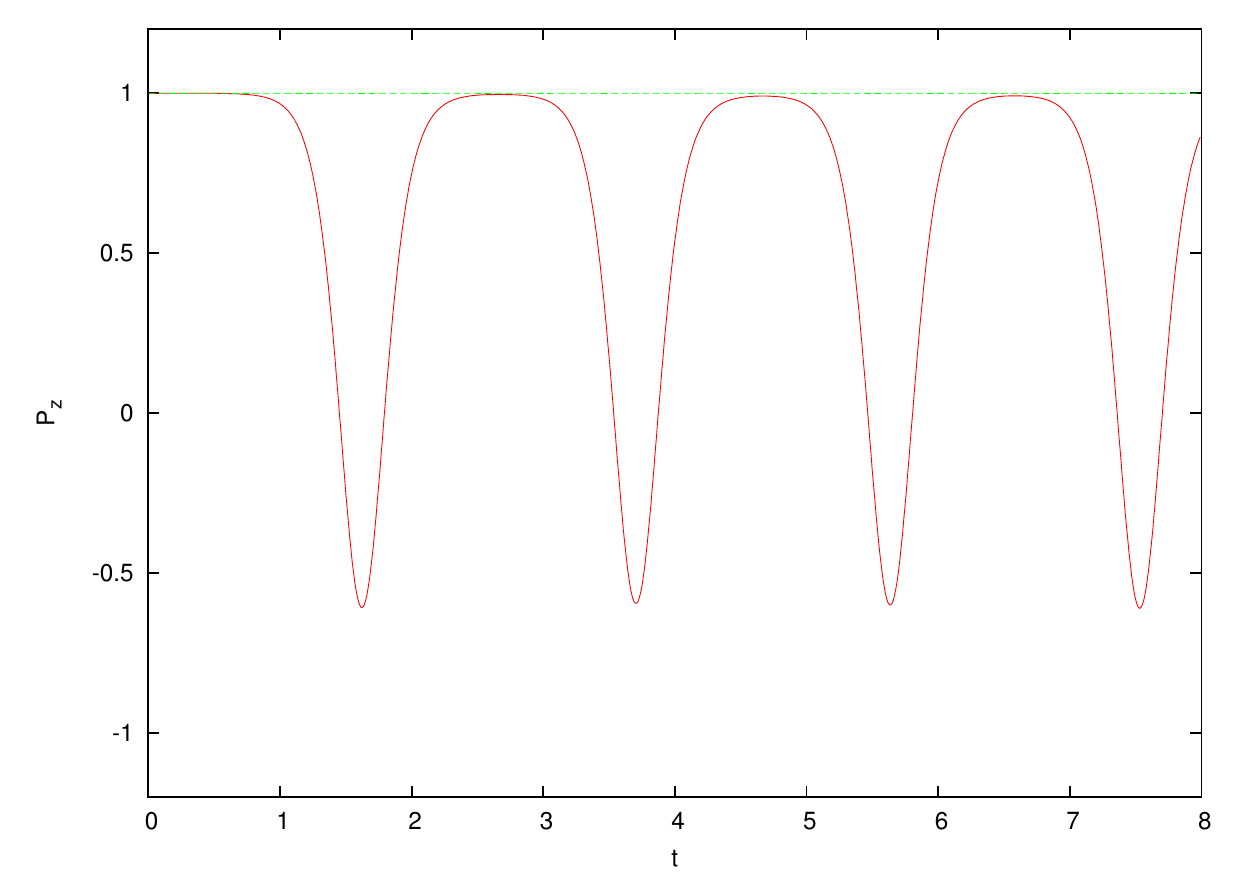}
\caption{Synchronized and Bipolar oscillation of neutrino. Solid line: Synchronized motion for $\mu=200$. Dashed Line: Bipolar motion for $\mu=10$. In this figure $\omega=1$, $\theta_{0}=10^{-2}$ and $\alpha=0.8$.}
\end{center}
\end{figure}
\begin{figure}
\includegraphics[scale=0.4]{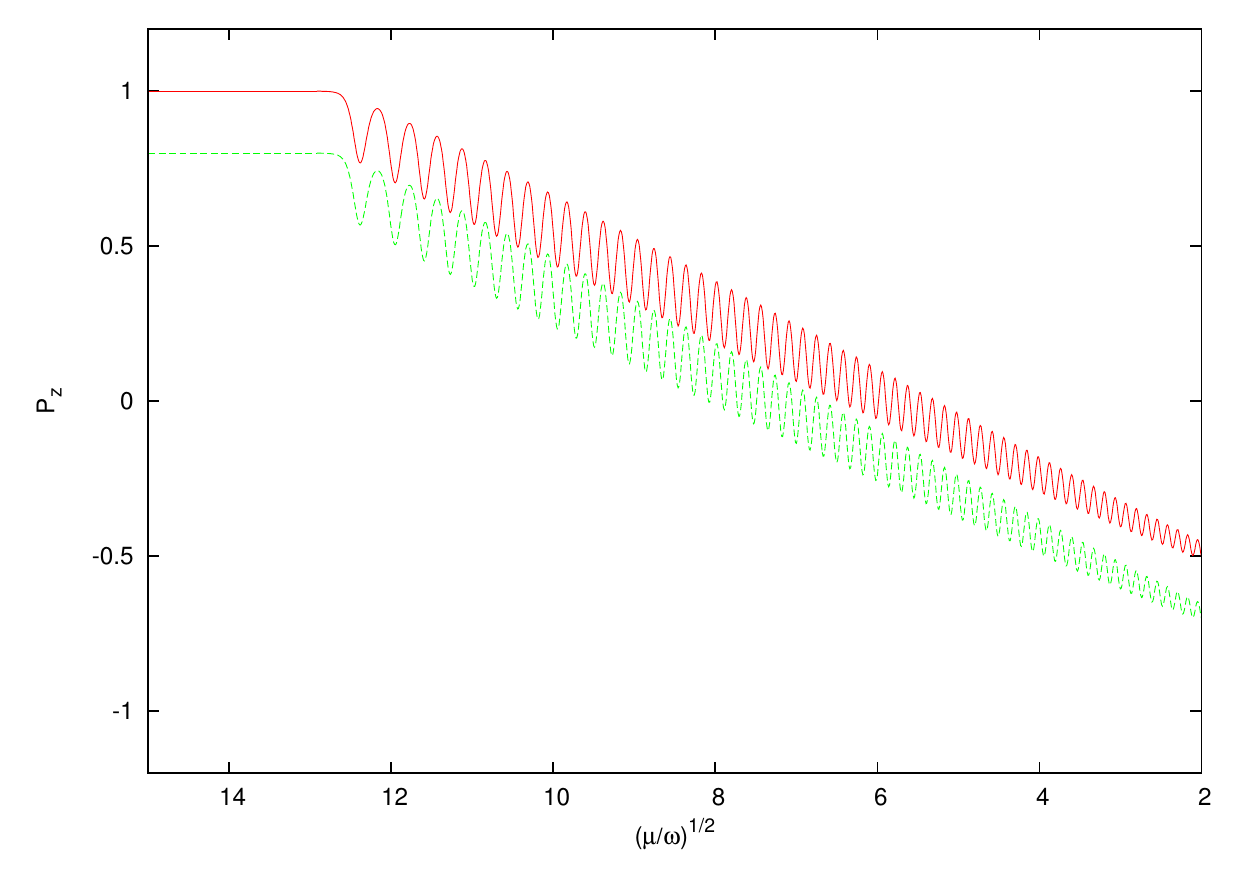}
\includegraphics[scale=0.4]{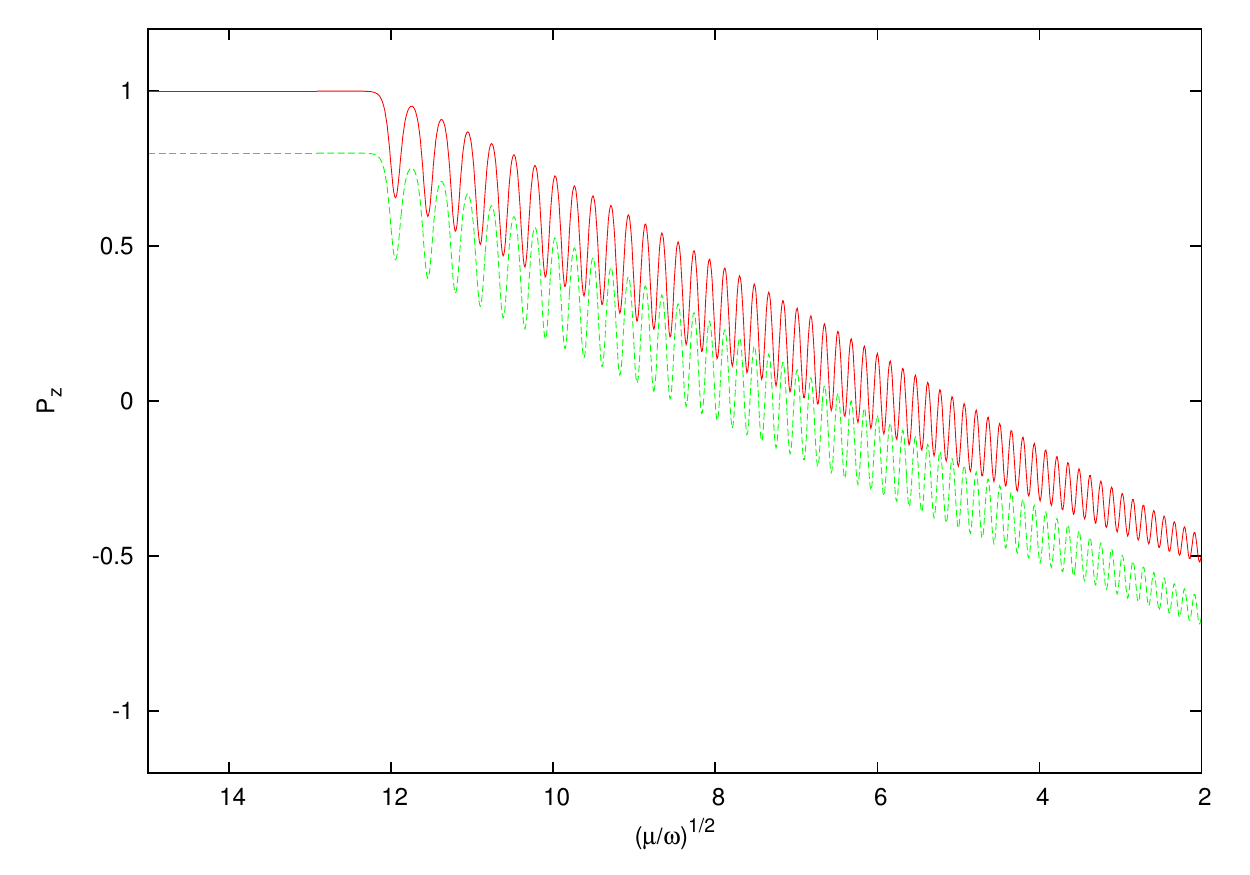}
\includegraphics[scale=0.4]{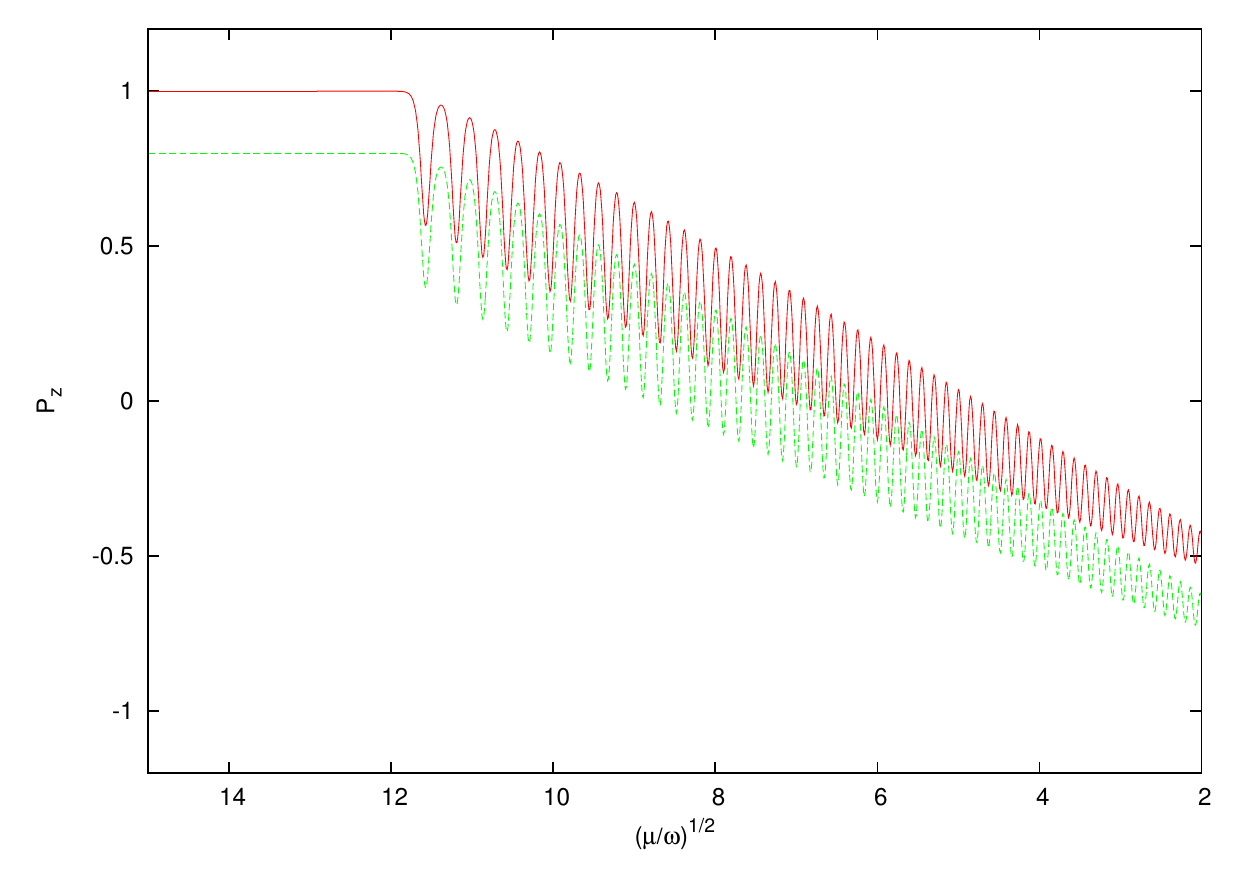}
\caption{Evolution of $P_{z}$ of neutrino (solid line) and $\bar{P}_{z}$ of antineutrino (dashed line) with decreasing $(\frac{\mu}{\omega})^{\frac{1}{2}}$, where $\theta_{0}=10^{-3}$ (left), $\theta_{0}=10^{-5}$ (middle) and $\theta_{0}=10^{-5}$ (right).}
\end{figure}
\begin{figure}
\begin{center}
\includegraphics[scale=0.7]{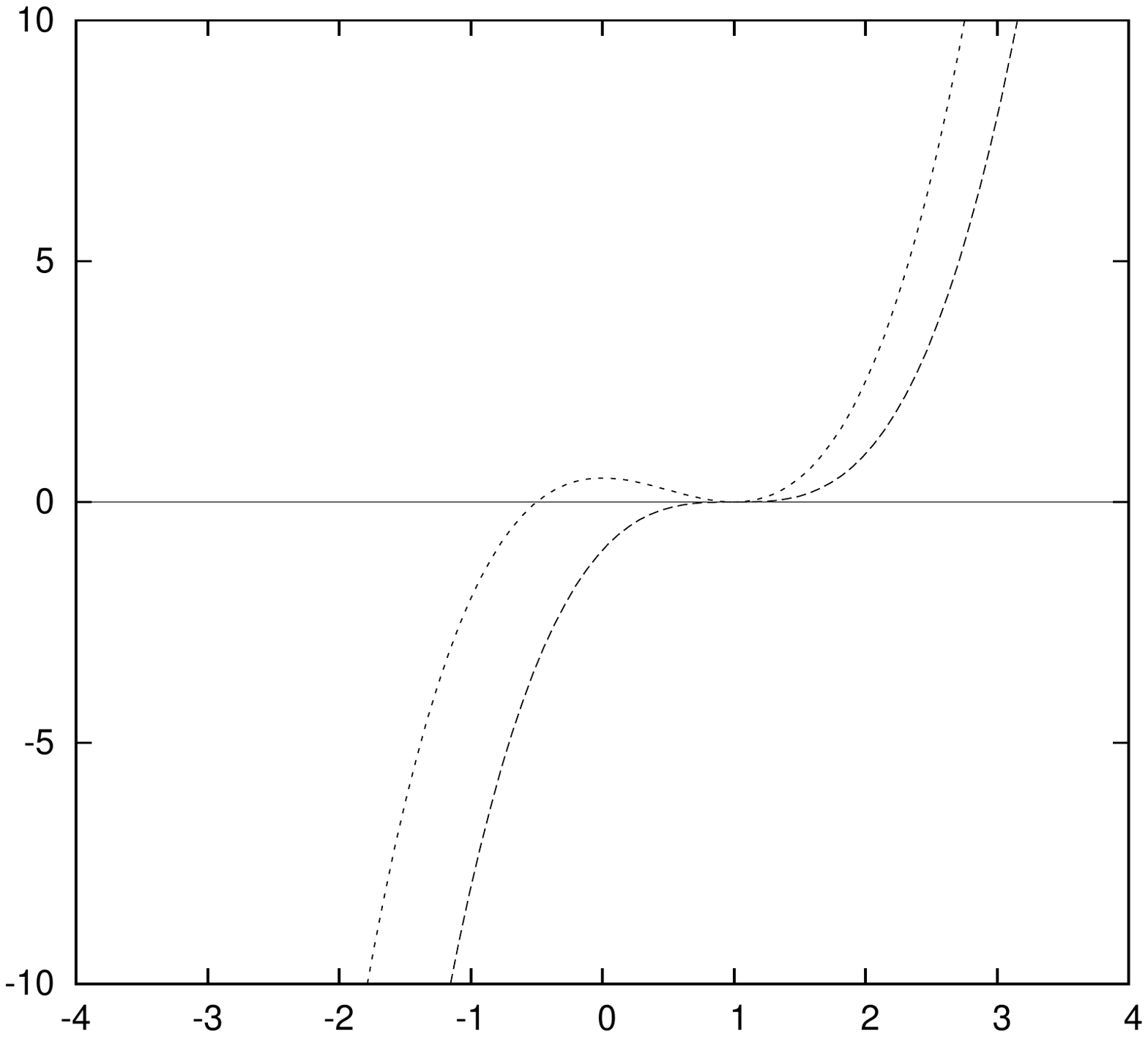}
\caption{The curve $f(u)$ of equation (13). Dark dashed line indicates the onset condition with double roots at $u_{0}=\cos 10^{-3}$, whereas light dashed line indicates there are two unequal roots between (-1,1).}
\end{center}
\end{figure}
\begin{figure}
\begin{center}
\includegraphics[scale=0.7]{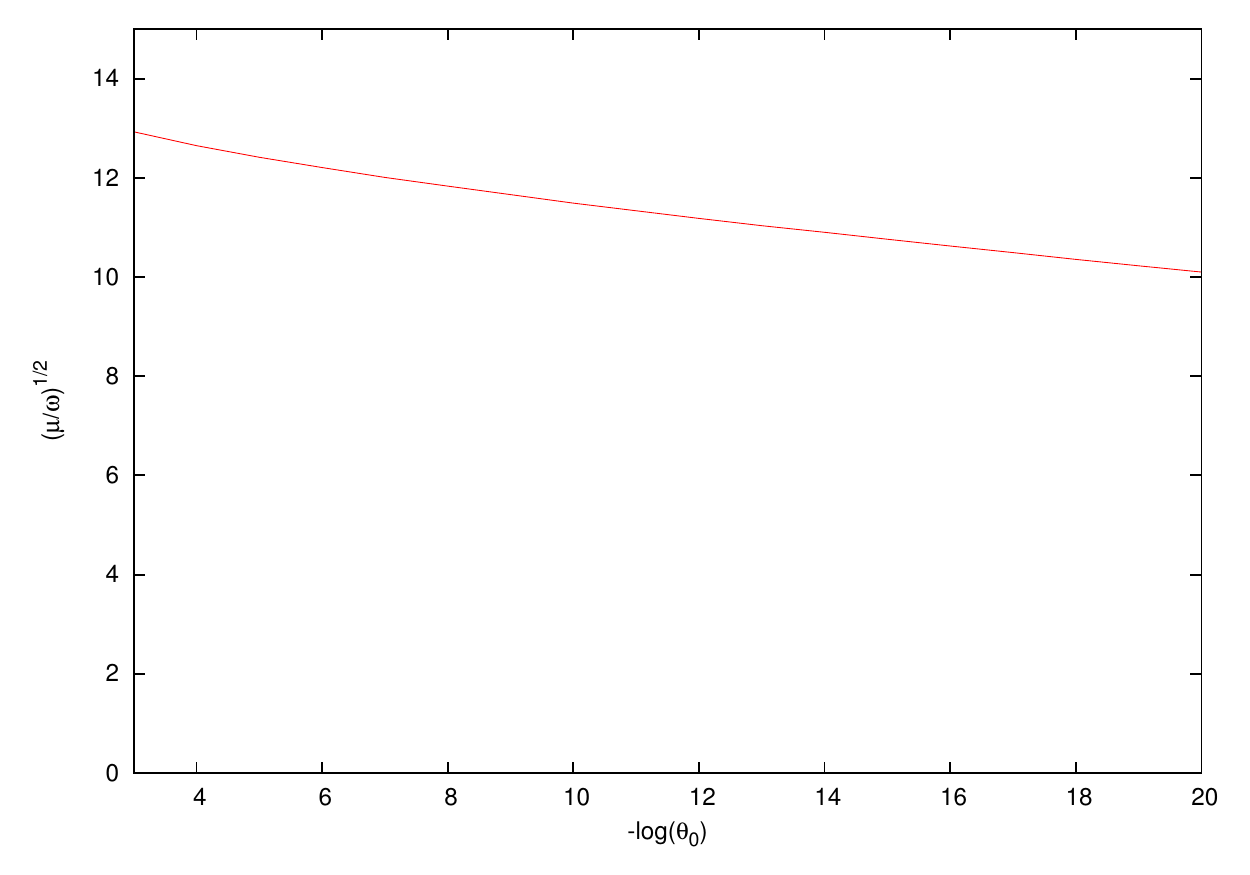}
\caption{Numerically obtained onset values of $(\frac{\mu}{\omega})^{1/2}$ against $-\log\theta_{0}$}
\end{center}
\end{figure}
\begin{figure}
\begin{center}
\includegraphics[scale=0.7]{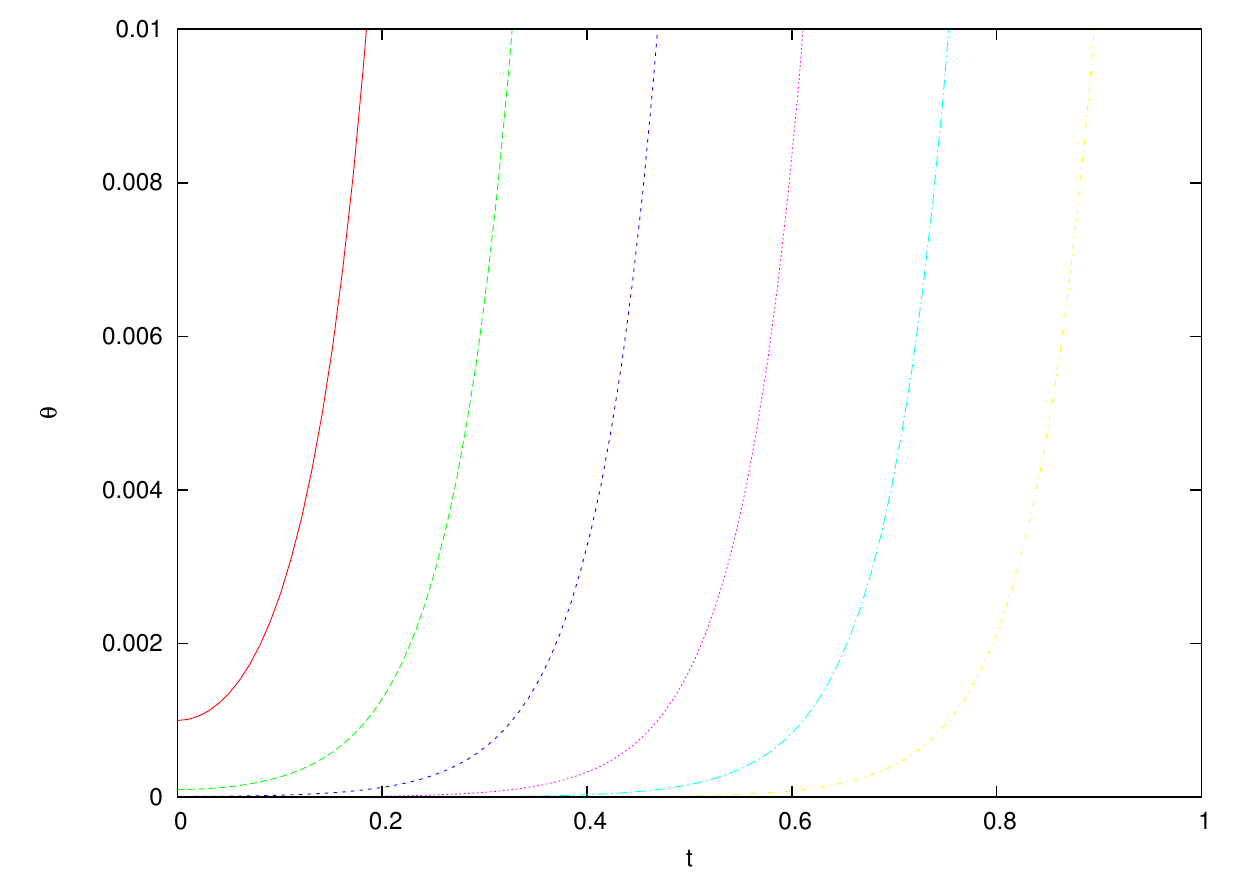}
\caption{The change of $\theta$ with time at $\theta_{0}$=$10^{-3}$ (red), $10^{-4}$ (green), $10^{-4}$ (blue), $10^{-5}$ (pink), $10^{-6}$ (light blue) and $10^{-7}$ (yellow).}
\end{center}
\end{figure}
\begin{figure}
\begin{center}
\includegraphics[scale=0.7]{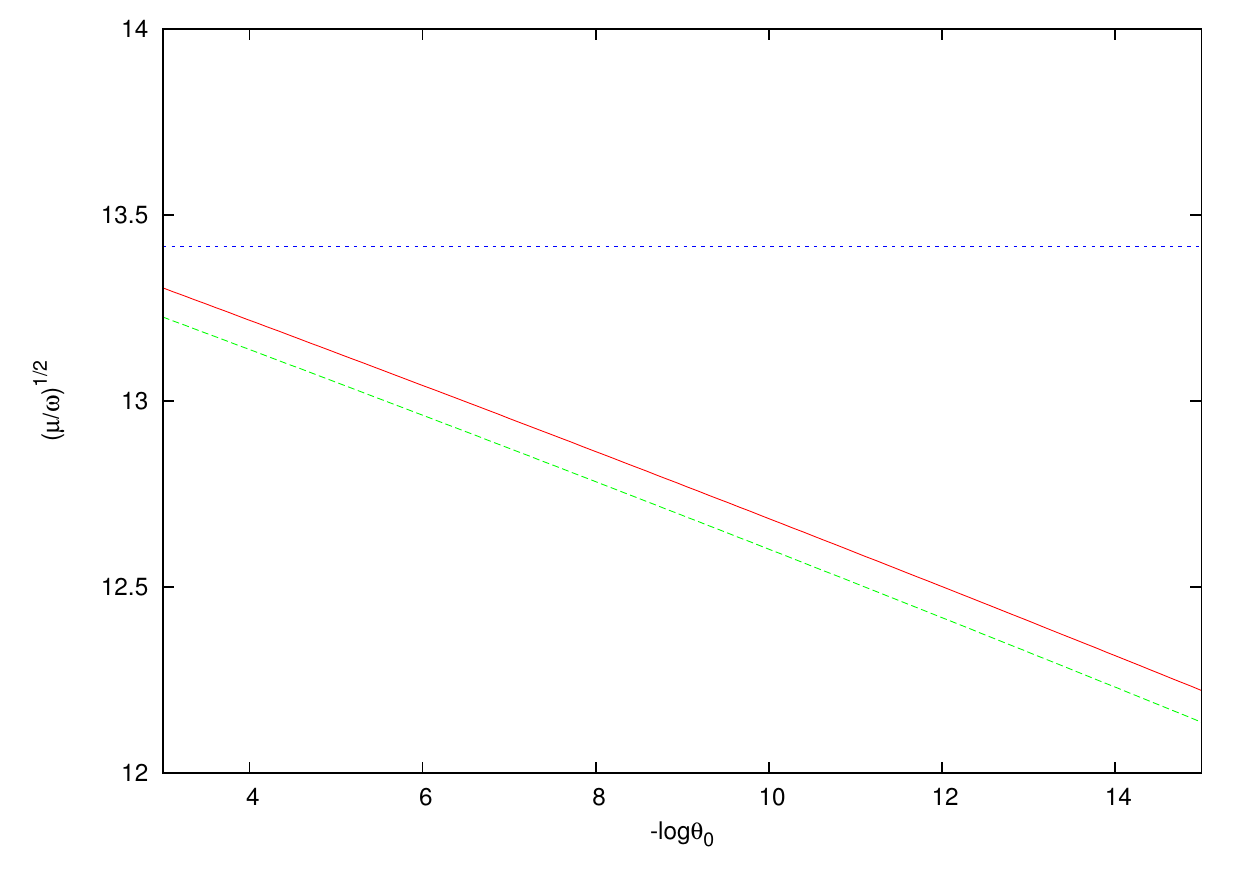}
\caption{The change of $(\frac{\mu}{\omega})^{1/2}$ with $\theta_{0}$ at $\theta=0.01$ (red) and $0.08$ (green). Here blue line indicates $\frac{\mu}{\omega}=180$, true onset value.}
\end{center}
\end{figure}
\begin{figure}
\begin{center}
\includegraphics[scale=0.7]{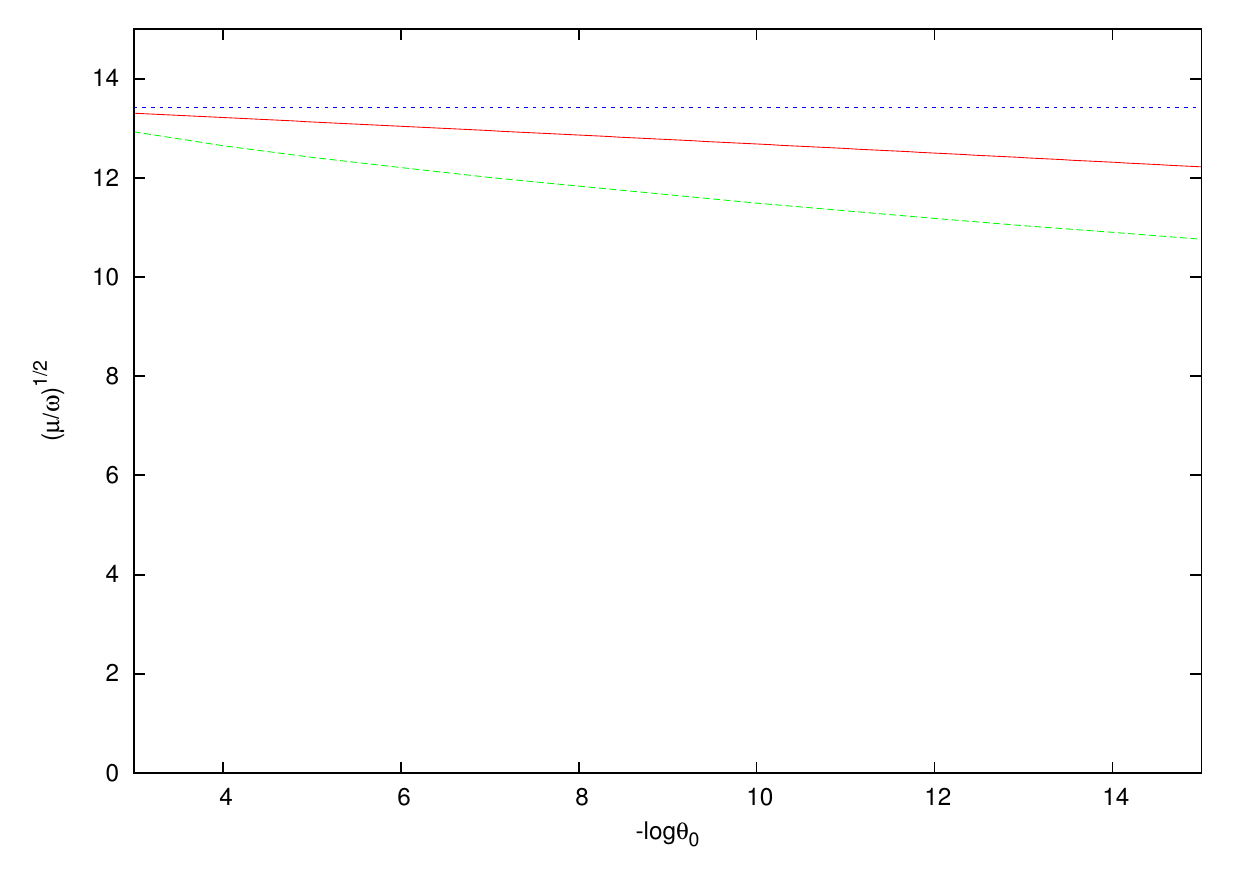}
\caption{Similar to the Figure-6; here the theoretical result at $\theta=0.01$ (red) is compared to the numerically obtained value (green).}
\end{center}
\end{figure}
\end{document}